\documentclass[twocolumn,showpacs,preprintnumbers,amsmath,amssymb,aps]{revtex4}
\usepackage{graphicx}
\begin{document}

\title{Realizing Colloidal Artificial Ice on Arrays of Optical Traps} 
\author{A. Lib{\' a}l,$^{1,2}$ C. Reichhardt,$^{1}$ 
 and 
C.J. Olson Reichhardt$^{1}$}
\affiliation{ 
{$^1$}Center for Nonlinear Studies and Theoretical 
Division, 
Los Alamos National Laboratory, Los Alamos, New Mexico 87545\\
{$^2$}Department of Physics, University of Notre Dame, Notre Dame, 
Indiana 46556
}

\date{\today}

\begin{abstract}
We demonstrate how a colloidal version of artificial ice 
can be realized on optical trap lattices. Using numerical simulations, we 
show that this system 
obeys the ice rules and that for strong colloid-colloid interactions, an 
ordered ground state appears.
We show that the ice rule ordering can occur for systems 
with as few as twenty-four traps and
that the ordering transition can be 
observed at constant temperature by varying the barrier 
strength of the traps.   
\end{abstract}
\maketitle

In certain spin models, the geometric spin arrangements frustrate 
the system since not all of the
nearest neighbor spin interaction energies can be minimized 
simultaneously \cite{ramirez94}.
A classic example of this is the spin ice 
system \cite{Harris97,BramwellBramwell01}, named after the 
similarity between magnetic ordering on a pyrochlore lattice and proton 
ordering in water ice \cite{AndersonPaulingKuo04}.
Spin ice behavior has been observed in magnetic materials such as 
Ho$_2$Ti$_2$O$_7$, where the magnetic rare-earth 
ions form a lattice of corner-sharing tetrahedra 
\cite{Harris97}.
The spin-spin interaction energy in 
such a system can be minimized locally when two spins in each tedrahedron 
point inward and two point outward, 
leading to exotic disordered states \cite{Ramirez}. There are several open 
issues in these systems, such as whether 
long range interactions order the system, or whether the true ground state 
of spin ice is ordered \cite{Melko01Siddharthan01}.

In atomic spin systems, the size scale is too small to examine ordering on 
the individual spin level directly, 
and very low temperatures are required to freeze the spins. Artificial 
versions of spin ice systems that overcome 
these limitations would be very useful. In a recent experiment, a 
geometrically frustrated system was constructed 
from a square lattice of small, single-domain magnetic islands \cite{Wang}. 
Each vertex of the lattice represents 
a meeting point for four spins. Wang {\it et al.} demonstrated that the 
system obeys the ``ice rules'' of
two-spins-in, two-spins-out at each vertex for closely spaced islands, and 
has a random spin arrangement for widely 
spaced islands. 
Unlike in atomic systems,
it is possible to image the ground state of the 
resulting spin ice directly using a scanning probe.

Here, we propose another version of an artificial spin ice system in which 
both statics and dynamics can be probed directly.
We use numerical simulations to show that square ice as well as other 
frustrated states can be constructed 
using interacting colloidal particles confined in two-dimensional (2D) 
periodic optical trap arrays. Due to the micron 
size scale of the colloids, the ordering and dynamics could be imaged with 
video-microscopy in an experiment. 
The colloidal system may also equilibrate much more rapidly than the 
nanomagnet system, since thermal fluctuations 
are present and can be controlled by changing the relative strength of the 
optical traps. In addition, 
the colloidal interaction can be changed from nearest neighbor to longer 
range simply by adjusting the screening length. 
A variety of different static and dynamical trap geometries can be 
constructed with optical arrays \cite{Grier}, 
and colloidal crystallization and melting have already 
been demonstrated in square and triangular optical trap 
arrays \cite{KordaReichhardtMangold,Bechinger}.
It is also possible to make arrays with elongated traps that have a 
double well shape such that a 
single colloid can be located in either well \cite{Babic,Bechinger2}. 
Colloid-colloid interaction forces between neighboring traps 
were strong enough to induce a zig-zag ordering 
and permit signal propagation in an experiment on a 
chain of 23 double well elongated traps
\cite{Babic}.

We consider an artificial ice system created with charged colloids on a 
square lattice of 
elongated optical traps
at a one-to-one filling.
Each trap has a double well potential similar to 
those created experimentally \cite{Babic,Bechinger} 
The vertices where four traps meet correspond to 
the oxygen atoms or the pyrochlore tetrahedrons, and the charged colloidal 
particles model the 'in' or 'out' spins. 
We classify the resulting six vertex types \cite{youngblood} according to 
their electrostatic energy, and study the change in the occupancy of 
different vertex types as a function of colloid charge and trap spacing. 
For noninteracting colloids we find random occupancy of each vertex type. 
When the colloid charge is increased, ice-rule obeying vertices dominate
the system. At high 
interaction strengths, we find a long-range ordered minimum energy state 
predicted previously for spin ice systems 
\cite{Melko01Siddharthan01}. We can recover the random vertex occupancy 
at high colloid charge by increasing
the spacing between traps. When the traps are gradually biased in one 
direction, we obtain a transition between two 
ice-rule obeying ground states. 
Since it can be difficult to construct very large arrays of traps 
experimentally, we show that a system with as few as 24 traps and
open boundary conditions still exhibits the 
ordering
transition observed for larger arrays. 
Previous experiments focused on order-disorder transitions for colloids
as a function of increasing trap strength \cite{Bechinger}.
Thus, we also demonstrate that the spin ordering transition occurs at constant
temperature when the trap barrier strength is varied.
  
We perform 2D Brownian dynamics (BD) simulations for systems of two sizes.
System A contains
$N=1800$ interacting colloids and $N=1800$ optical traps with periodic 
boundary conditions in the $x$ and $y$ directions. 
System B has $N=24$ colloids and $N=24$ optical traps with open
boundary conditions.
In each case the overdamped equation of motion for colloid $i$ is:
\begin{equation}
\eta\frac{ d{\bf R}_{i}}{dt} = {\bf F}_{i}^{cc} + {\bf F}^{T}_{i} + 
{\bf F}^{ext}_{i} + {\bf F}^{s}_{i} 
\end{equation}
where the damping constant $\eta=1.0$. 
We define the unit of distance in the simulation to be $a_0$.
The colloid-colloid interaction force has a Yukawa or 
screened Coulomb form, 
${\bf F}_{i}^{cc} = -F_0q^2\sum^{N}_{i\neq j}\nabla_i V(r_{ij})$
with
$V(r_{ij}) = (1/r_{ij})\exp(-\kappa r_{ij}){\bf {\hat r}}_{ij}$.
Here $r_{ij}=|{\bf r}_{i} - {\bf r}_{j}|$,
${\bf {\hat r}}_{ij}=({\bf r}_{i}-{\bf r}_{j})/r_{ij}$,
${\bf r}_{i(j)}$ is the position of particle $i$($j$),
$F_0=Z^{*2}/(4\pi\epsilon\epsilon_0)$,
$Z^*$ is the unit of charge, $\epsilon$ is the 
solvent dielectric constant,
$q$ is the dimensionless colloid charge,
and $1/\kappa$
is the screening length, 
where $\kappa=4/a_0$ unless
otherwise mentioned.
We neglect hydrodynamic interactions between colloids, which is a reasonable
assumption for charged particles in the low volume fraction limit.
The thermal force ${\bf F}^T$ is modeled as random Langevin kicks with the 
properties $\langle{\bf F}^{T}_{i}\rangle = 0$ and
$\langle {\bf F}^{T}(t) {\bf F}^{T}(t^{\prime})\rangle = 
2\eta k_{B}T\delta(t - t^{\prime})$. 
Unless otherwise mentioned, $F^T=|{\bf F}^{T}|=0$.
${\bf F}^{ext}_{i}$ represents an externally applied
drive which is set to zero except for the biased system, where 
${\bf F}^{ext}_{i} = F^{dc}({\bf {\hat x}} + {\bf {\hat y}})$.  

\begin{figure}
\includegraphics[width=3.3in]{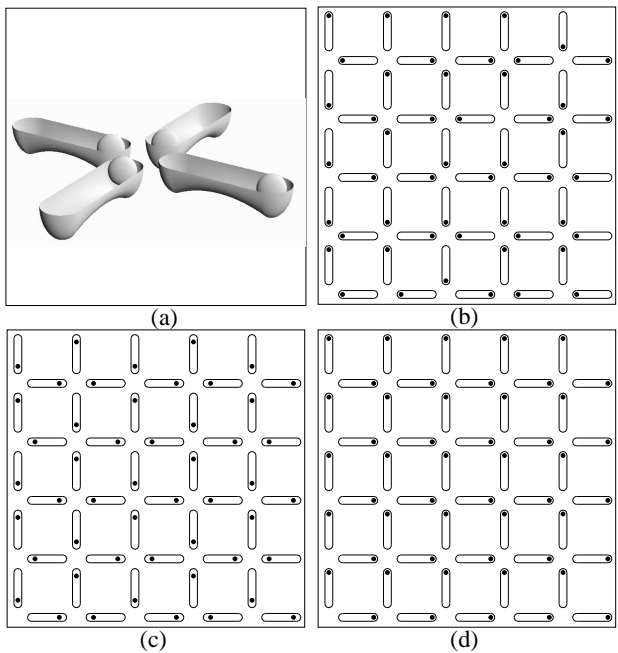}
\caption{
(a) Schematic of the basic unit cell with four double well 
traps each capturing one colloid.   
(b-d) Images of a small portion of system A with $N=1800$.
Dark circles: colloids; ellipses: traps.
(b) Random vertex distribution at $q=0$.
(c) Long-range ordered square ice ground state at $q=1.3$.
(d) Biased system at $q=0.4$ with $F^{dc}=0.02$.
}
\end{figure}

\begin{table}
\begin{tabular}{|c|c|c||c|c|c|}
\hline
Type & Configuration & $E_i/E_{III}$ & Type & Configuration & $E_i/E_{III}$\\
\hline
I & 0000 & 0.001 & IV & 1001 & 7.02\\
II & 0001 & 0.0214 & V & 1101 & 14.977\\
III & 0101 & 1.0 & IV & 1111 & 29.913\\
\hline
\end{tabular}
\caption{Electrostatic energy $E_i/E_{III}$ 
for
each vertex type. 
An example configuration for each vertex is
listed; 1 (0) indicates a colloid close to (far from) the vertex. 
}
\end{table}
 
The substrate force ${\bf F}^{s}_{i}$ arises from elongated traps, shown 
schematically in Fig.~1(a), arranged in square 
structures with lattice constant $d$, as in Fig.~1(b).
Each trap is composed of two half-parabolic 
wells of strength $f_p$ and radius $r_p$ 
separated by an elongated region of length $2l$ which confines the 
colloid perpendicular 
to the trap axis and has a small repulsive potential or barrier
of strength $f_r$ 
parallel 
to the axis which pushes the colloid out of the middle of the trap into 
one of the ends:
${\bf F}^{s}_{ik}=
(f_p/r_p)r_{ik}^\pm \Theta(r_p-r_{ik}^\pm){\bf \hat{r}}_{ik}^\pm
+(f_p/r_p)r_{ik}^\perp \Theta(r_p-r_{ik}^\perp){\bf \hat{r}}_{ik}^\perp
+(f_r/l)(1-r_{ik}^\parallel) \Theta(l-r_{ik}^\parallel)
{\bf \hat{r}}_{ik}^\parallel$.
Here
$r_{ik}^\pm=|{\bf r}_i-{\bf r}_k^p \pm l{\bf {\hat p}}^k_\parallel|$,
$r_{ik}^{\perp,\parallel}=
|({\bf r}_i - {\bf r}_k^p) \cdot {\bf {\hat p}}^k_{\perp,\parallel}|$,
${\bf r}_i$  (${\bf r}_k^p$)
is the position of colloid $i$ (trap $k$),
and ${\bf \hat{p}}^k_\parallel$ (${\bf \hat{p}}^k_\perp$) is a unit
vector parallel (perpendicular) to the axis of trap $k$.
We take $2l=2a_0$, $r_p=0.4a_{0}$, 
and $d=3a_0$ 
unless otherwise noted. 
Elongated traps of this form have been created in
previous experimental work \cite{Babic,Bechinger2}.
Our dimensionless units
can be converted to physical units for a particular system.  For example, 
when $a_0=2$ $\mu$m, $\epsilon=2$, 
and $Z^*=300e$, such as in Ref.~\cite{Dufresne},
$F_0=2.5$ pN and the trap ends are 0.2 $\mu$m apart at $d=3$. 
We find the ground state of each configuration using simulated annealing. 

The vertices are categorized into six types, listed in Table I, and we 
identify the percentage occupancy $N_i/N$ and energy $E_i$
of each type.
Type III and type IV vertices each obey the ice rule 
of a two-in two-out configuration, represented here by two colloids 
close to the vertex and two far from the vertex. 
Locally, the system would prefer type I vertices, but such vertices must 
be compensated by highly unfavorable type 
VI vertices. 
The colloidal spin ice realization differs from the 
magnetic system, where north-north 
and south-south magnetic interactions at a vertex have equal energy. For 
the colloids, interactions between two 
filled trap ends raise the vertex energy $E_i$, whereas two adjacent empty 
trap ends decrease $E_i$.
Since particle number must be conserved, creating empty trap ends at one 
vertex increases the particle load at 
neighboring vertices. As a result, the ice rules still apply to our system, 
but they arise due to collective 
effects rather than from a local energy minimization. 

In Fig.~1(b) we illustrate a small part of 
system A with noninteracting colloids at charge $q=0$.
The distribution of $N_i/N$ is consistent with a random arrangement.
When we increase 
$q$ to $q=1.3$ so that the colloids are strongly interacting, we find a 
nonrandom configuration 
where the system is filled with type III vertices in a checkerboard pattern 
corresponding to the 
square ice ground state \cite{Stillinger}, illustrated in Fig. 1(c).  We 
find similar behavior 
in system B containing only 24 traps,
which may be easier to realize in an experiment.

\begin{figure}
\includegraphics[width=3.35in]{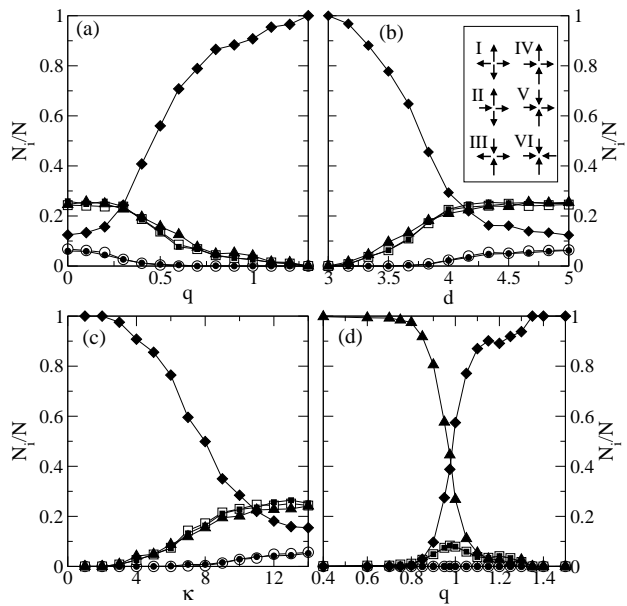}
\caption{
$\bigcirc$: $N_I/N$; $\square$ : $N_{II}/N$; $\blacklozenge$: $N_{III}/N$;
$\blacktriangle$: $N_{IV}/N$; $\blacksquare$: $N_{V}/N$; $\bullet$:
$N_{VI}/N$.
(a) $N_i/N$ vs $q$ at $d=3$ and $\kappa=4.0$.
(b) 
$N_i/N$ vs $d$ at $q=1.3$ and $\kappa=4.0$.
Inset: schematic spin representation of the 6 vertex types.
(c) $N_i/N$ vs $\kappa$ at $d=3$ and $q=1.0.$
(d) $N_i/N$ vs $q$ for a biased system at $d=3$, $\kappa=4.0$,
and $F^{dc}=0.02$.
}
\end{figure}

We use $N_i/N$ to map the transition between the random state and the 
long-range ordered
state as a function of 
colloid charge 
$q$ at fixed trap spacing 
$d=3$ and $\kappa=4.0$ in Fig.~2(a), 
as a function of $d$ for fixed $q=1.3$ and $\kappa=4.0$ 
in Fig.~2(b), and as a function of $\kappa$ for fixed
$q=1.0$ and $d=3$ in Fig.~2(c).
Changing $q$, $d$ or $\kappa$ 
changes the relative colloid-colloid interaction
strength. In Fig.~2(a), 
$N_I/N$ and $N_{VI}/N$ 
decrease with increasing $q$ for $q>0.1$, since this is the most energetically
unfavorable vertex combination. As $q$ 
increases above $q>0.3$, $N_{II}/N$, $N_{V}/N$ and 
$N_{IV}$ begin to decrease since type II, V and IV vertices are 
also energetically 
unfavorable. 
Even though type IV vertices obey ice rules, they disappear at high $q$
since in square ice 
they have higher energy than the type III vertices, 
which
ultimately form the ground state.

In Fig.~2(b) we fix $q=1.3$ and $\kappa=4.0$ and show that changing the 
trap spacing $d$ from $d=3$ to $d=5$
produces the same transition as changing the colloid charge. 
As 
$d$ increases, the 
colloid-colloid interaction strength
drops, and the vertices that were suppressed by the high colloid charge $q$
reappear in the same order: types II, V and IV first, 
followed by types I and VI.
If $q$ and $d$ are fixed and the inverse screening length $\kappa$ is
varied, we find that a similar transition from an ordered to a random
configuration occurs, as illustrated in Fig.~2(c) 
for $d=3$ and $q=1.0$.
In experiments with dynamic optical traps, 
it would be straightforward to adjust the trap spacing $d$, 
allowing both
random and ordered limits to be accessed readily. 
The appearance of an ordered state for small trap spacing is similar to
experimental observations of spin liquid ordering under pressure \cite{Mirebean02}.

From these results we see that a colloidal system can serve as a model for 
artificial ice. The advantage of the colloid realization is that it is 
possible to 
directly observe the dynamical behavior of the individual ``spins.''  It is 
also
possible to create more complex systems than can be achieved with the 
magnetic system. 
For instance,
to create biased traps, we introduce a small driving force oriented at
a $45^{\circ}$ angle along the diagonal of the square plaquettes.  This
causes the colloids to favor sitting in the topmost and rightmost halves 
of the traps, and breaks symmetry in the same way as applying a
magnetic field along the [100] direction in a pyrochlore system
\cite{harris98,ramirez99,fukayawa02,hiori03}.
In Fig.~1(d) we illustrate the ground state of a system biased in this way
with $q=0.4$, $d=3$, $\kappa=4$, and $F^{dc}=0.02$. 
All of the traps are effectively tilted in the rightward and upward 
directions, and the system has a ground state made up entirely 
of type IV vertices. 
In Fig.~2(d), we see that in the biased system, 
increasing $q$ produces a transition 
between the tilted and the spin ice ground states. 
At the transition, we find vertex clusters composed entirely of type III or
type IV vertices, with pairs of type II and type V vertices located at
the boundaries between the clusters.  Type I and type VI vertices never
appear in this system.

\begin{figure}
\includegraphics[width=3.25in]{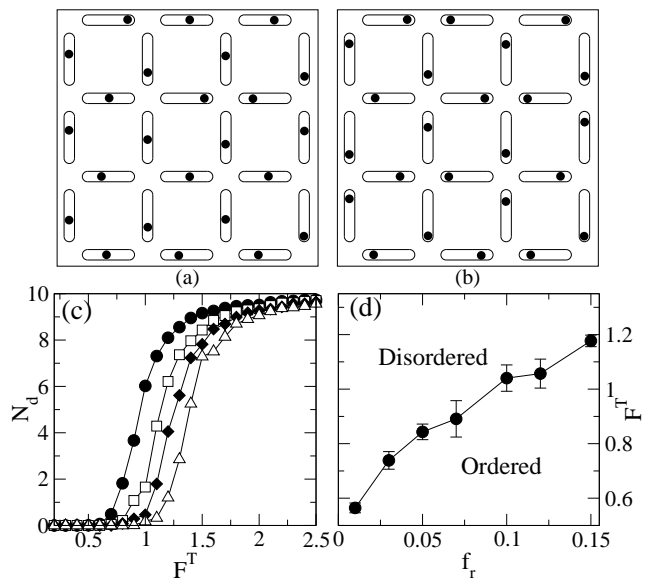}
\caption{
(a,b) Images of the entire sample for system B, 
with $N=24$ and open boundary conditions at $d=3$, $q=1$, $\kappa=4.0$,
and $F^T=0.9$. (a) Disordered system at $f_r=0.01$.
(b) Square ice ground state at $f_r=0.10$.
(c) The number of defects $N_d$ versus temperature $F^T$
for the same system with $f_r=0.03$ (filled circles), $f_r=0.07$ 
(open squares),
$f_r=0.1$ (filled diamonds), and $f_r=0.15$ (open triangles). 
(d) A phase diagram showing the ordered and disordered states
as a function of $F^T$ vs 
$f_r$.  
}
\end{figure}

In experimental systems using current equipment, a limited number of optical
traps are available.  For example, 23 traps were employed in
recent experiments on double well elongated traps \cite{Babic}.
It is therefore pertinent to determine how small of a system 
can still exhibit the artificial ice behavior. 
Additionally, experimental systems  have no periodic
boundary conditions, so it is important to understand whether the
boundaries affect the response of the system.

In Fig.~3(a,b) we illustrate two configurations of system B, which contains
24 traps and has open boundary conditions, for the case 
$q=1.0$, $d=3.0$, and $\kappa=4.0$ at a temperature $F^T=0.9$.
A disordered configuration, shown in Fig.~3(a), results when the
strength of the repulsive barrier at the center of each trap, $f_r=0.01$, 
is small.
Here the thermal effects are strong
enough that the colloids are hopping between the two different wells
in each trap. 
As $f_{r}$ is increased, a freezing transition occurs and the system
forms the ordered phase, seen in Fig.~3(b) for $f_r=0.10$. 
To quantify the ordering transition, we measure the time-averaged number of
defects in the system, $N_d$, which is determined by comparing the
colloid configuration $C$ with the two possible ground states for the 24
traps: $G$, shown in Fig.~3(b), and $G^\prime$, obtained by flipping
each colloid in Fig.~3(b) to the opposite side of each trap.
We take 
$N_d=\langle \min(\sum_i^N{|C_i-G_i|},\sum_i^N{|C_i-G^\prime_i|})\rangle$,
where $|C_i-G_i^{(\prime)}|=0$ (1) if colloid $i$ is at the same (opposite) 
end of the trap in the two configurations.
A plot of $N_d$ versus $F^T$ in Fig.~3(c) shows that
as $F^T$ is decreased, there is a freezing transition into the ordered state
where $N_d=0$. The freezing temperature decreases as $f_r$ is lowered.
The most straightforward experiment to perform would be to fix temperature
and the other parameters while varying the strength of the barrier at the
center of the traps, $f_r$.
To illustrate that the order-disorder artificial
ice transition can also occur as function of barrier strength, we
map out a phase diagram as a function of $f_r$ and $F^T$ in Fig.~3(d). 
This shows that for systems as small as 24 traps, the artificial ice
behavior should be experimentally observable.

All of the systems we have studied up to this point have a well 
defined, long-range ordered ground state. 
We have also found that if a honeycomb arrangement of traps is used
instead of the square trap arrangement considered here,
only disordered ground states
occur \cite{Libal2}.   

In conclusion,
we have shown that an artificial ice model system can be created using charged 
colloidal particles in arrays of elongated optical traps. 
The system obeys the ice rules and shows a transition between 
a random configuration and a long-range ordered ground state as a
function of colloid charge, trap size, and
screening length.  
We demonstrate that a thermally induced order-disorder
transition also occurs in samples with only 24 traps
and open boundary conditions, which should be well within
the range of current experimental capabilities.
This transition can be observed at fixed temperature by
varying the trap barrier strength, which would be the most straightforward
experiment to conduct.
Besides optical traps, other systems including electrophoretic traps
\cite{Cohen05}
or patterned surfaces may also be used to confine the colloids. 
Similar effects should occur for
vortices in type-II superconductors interacting with
elongated arrays of blind holes. 
Experimental versions
of frustrated colloidal systems could allow for direct 
visualization of the dynamics associated with
frustrated spin systems, such as deconfined or
confined spin arrangements, as well as spin dynamics at
melting transitions.

{\it Note added-}
A simulation study \cite{Moessner}
of the experimental dipolar system appeared after submission of this work.
 
This work was supported by the US Department of Energy under Contract No.
W-7405-ENG-36.

\end{document}